\documentclass[twocolumn, amsmath,amssymb, superscriptaddress]{revtex4}

\usepackage{dcolumn}
\usepackage{bm}
\usepackage{amsmath}
\usepackage{amsfonts}
\usepackage{graphics}
\usepackage[pdftex]{graphicx}
\usepackage{times}
\usepackage{setspace}
\usepackage{verbatim}
\usepackage{enumerate}
\usepackage{hyperref}
\hypersetup{colorlinks=true,citecolor=blue, linkcolor=black,urlcolor=blue}
\usepackage{color}   
\usepackage{framed}
\usepackage[raggedright]{titlesec}

\begin{document}

\title{Reputation and Impact in Academic Careers}
 
\author{Alexander M. Petersen}
\affiliation{Laboratory for the Analysis of Complex Economic Systems, IMT   Institute for Advanced Studies Lucca,  55100 Lucca, Italy}
\author{Santo Fortunato}
\affiliation{Department of Biomedical Engineering and Computational Science, Aalto University School of Science, P.O. Box 12200, FI-00076, Finland}
\author{Raj K. Pan}
\affiliation{Department of Biomedical Engineering and Computational Science, Aalto University School of Science, P.O. Box 12200, FI-00076, Finland}
\author{Kimmo Kaski}
\affiliation{Department of Biomedical Engineering and Computational Science, Aalto University School of Science, P.O. Box 12200, FI-00076, Finland}
\author{Orion Penner}
\affiliation{Laboratory of Innovation Management and Economics, IMT  Institute for Advanced Studies Lucca,  55100 Lucca, Italy}
\author{Armando Rungi}
\affiliation{Laboratory for the Analysis of Complex Economic Systems, IMT   Institute for Advanced Studies Lucca,  55100 Lucca, Italy}
\author{Massimo Riccaboni}
\affiliation{Laboratory of Innovation Management and Economics, IMT  Institute for Advanced Studies Lucca,  55100 Lucca, Italy}
\affiliation{Department of Managerial Economics, Strategy and Innovation, Katholieke Universiteit Leuven, 3000 Leuven, Belgium}
\author{H. Eugene Stanley}
\affiliation{Center for Polymer Studies and Department of Physics, Boston University, Boston,
Massachusetts 02215, USA}
\author{Fabio Pammolli}
\affiliation{Laboratory for the Analysis of Complex Economic Systems, IMT Institute for Advanced Studies Lucca,  55100 Lucca, Italy}
\affiliation{Center for Polymer Studies and Department of Physics, Boston University, Boston, Massachusetts 02215, USA}
\date{\today}

\begin{abstract} 
Reputation is an important social construct in science, which enables  informed  quality assessments  of both publications and careers of scientists in the absence of complete systemic information. However, the relation between reputation and career growth of an individual remains poorly understood, despite recent proliferation of quantitative research evaluation methods. Here we develop an original framework for measuring how a publication's citation rate $\Delta c$ depends on the reputation of its central author $i$, in addition to its net citation count $c$. To estimate the strength of the reputation effect, we perform a longitudinal analysis on the careers of 450 highly-cited scientists, using the total  citations $C_{i}$ of each scientist as his/her reputation measure. We find a citation crossover $c_{\times}$ which distinguishes  the strength of the reputation effect. For publications with  $c < c_{\times}$, the  author's reputation is found to dominate the annual citation rate.  Hence,  a new publication may gain a significant  early advantage corresponding to roughly a 66\% increase in the citation rate for each tenfold increase  in $C_{i}$. However, the reputation effect becomes negligible for highly cited publications meaning that for  $c\geq c_{\times}$ the citation rate  measures scientific impact more transparently. In addition we have developed a stochastic reputation model, which is found to reproduce numerous statistical observations for real careers, thus providing insight into the microscopic mechanisms underlying cumulative advantage in science.
\end{abstract} 

\maketitle 

\footnotetext[1]{Published in: Proceedings of the National Academy of Science USA, 2014. \href{http://www.pnas.org/content/early/2014/10/03/1323111111.abstract}{DOI:10.1073/pnas.1323111111}
Send correspondence to:\\ petersen.xander@gmail.com, santo.fortunato@aalto.fi, or hes@bu.edu}
Citation counts are widely used to judge the impact of both scientists and their publications~\cite{UnivCite,DiffusionRanking,Scientists,RankCitSciRep}. While it is recognized that many factors outside the pure merit of the research or the authors influence such counts, little effort has been devoted to identifying and quantifying the role of the author specific factors. Recent investigations  have begun to study the impact the individual scientists have through  collaboration and reputation spillovers~\cite{StarDeath,GrowthCareers},  two integrative features of scientific careers that contribute to cumulative advantage~\cite{Matthew1,cumadvprocess,BB2}. However, the majority of citation models avoid author specific effects, mainly due to the difficulty in acquiring comprehensive disambiguated career data~\cite{PetersonPNAS,PrefAttachModel,nonlinprefattach3,nonlinprefattach2}.  As the 
measures are becoming increasingly common in evaluation scenarios throughout science, it is crucial to better understand 
what  the citation measures actually represent in the context of scientists' careers. Moreover,  how does reputation affect a scientist's    access to key resources, the incentives to publish quality over quantity,  and other key decisions along the career path  
\cite{StephanJEL,EconScience,Caution1,Predictability2,ResourcesScience}? And what role does reputation  play in  the ``mentor matching'' process within academic institutions,  in the effectiveness of single/double blinding in peer-review, and  in  the reward system of science  \cite{StephanJEL,EconScience,OpenScience}?

\begin{figure}
\centering{\includegraphics[width=0.5\textwidth]{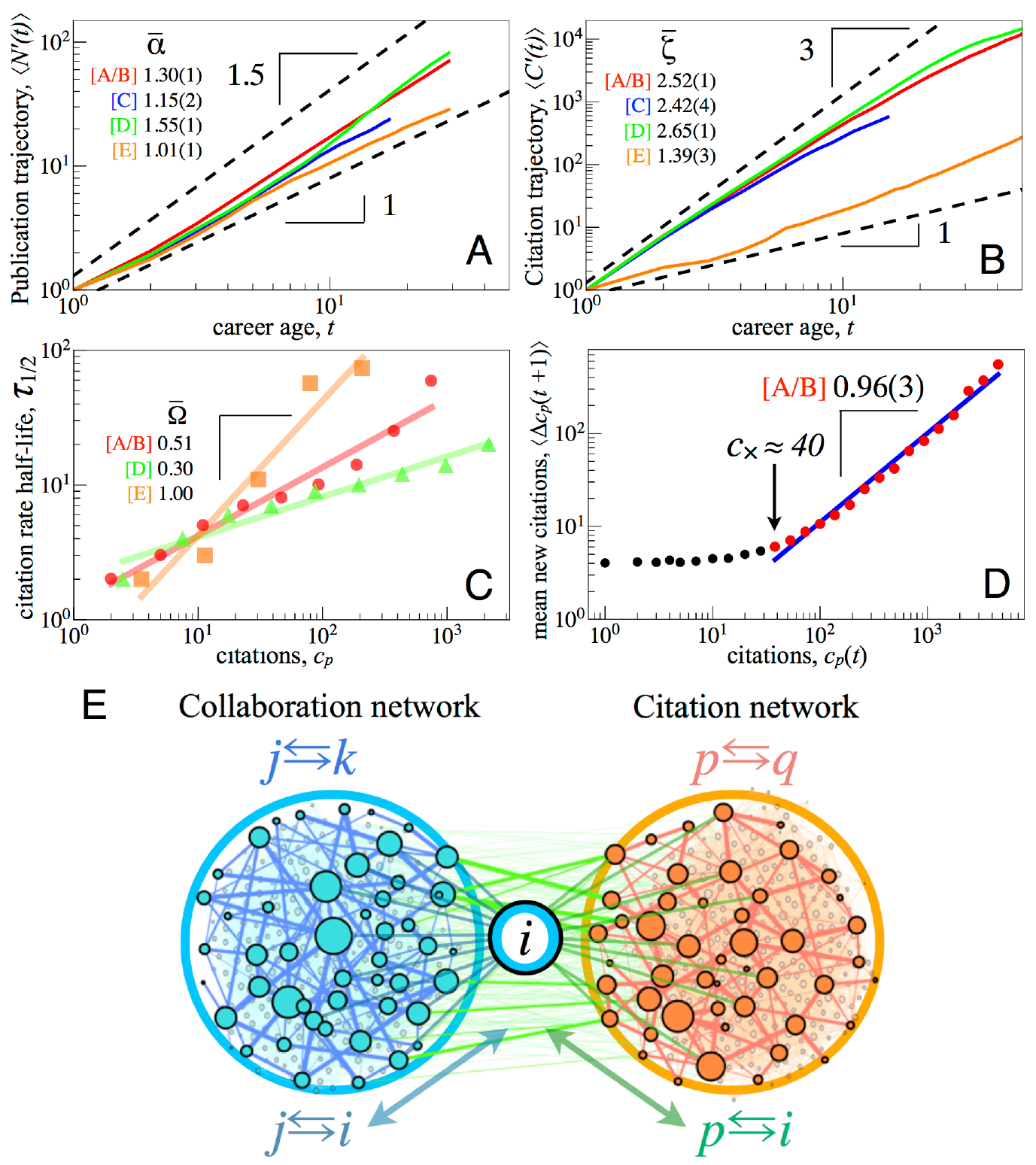}}
\caption{\label{NtCt}  Quantifying  cumulative reputation measures and citation dynamics.
 (A,B)  Growth trajectories of the cumulative publications $N'(t)$ and
    citations $C'(t)$, appropriately
  rescaled to start from unity in each ordinate. The characteristic
  $\overline \alpha$ and $\overline \zeta$ exponents shown in each
  legend are calculated over the growth phase of the career. The mathematicians [E] have distinct career trajectories, with
  $\overline \alpha \approx 1$ since collaboration spillovers via division of labor likely play a
  smaller role in publication rate growth. See Tables S1--S9 for  $\alpha_{i}$ and $\zeta_{i}$ values calculated for individual careers.
  ({C})  Relation between $\tau_{1/2}$ and cumulative citations $c_{p}$.
  (D)  Preferential attachment dynamics with $\pi \approx 1$ break down for $c < c_{\times}$. The reputation effect provides a citation boost above the baseline preferential attachment citation rate attributable to $c_{p}(t)$ only.
  }
\end{figure}

It is against this background that we have developed a quantitative framework with the goal of isolating the effect of author reputation upon citation dynamics. Specifically, by controlling for time- and author- specific factors, we quantify the role of author reputation
 on the citation life cycle of individual publications at the micro level. We use a longitudinal career dataset from Thomson Reuters Web of Science comprising 450 highly-cited scientists, 
83,693 articles and 7,577,084 citations tracked over 387,103 publication
years. 
Dataset [A] refers to 100 top-cited physicists, [B] to
another set of 100 highly prolific physicists, [C] to 100 
assistant professors in physics, [D] to 100 top-cited cell biologists,
and [E] to 50 top-cited pure mathematicians (for further data elaboration see the Supporting Information (SI) Appendix). 
For each central scientist $i$ we analyze the scientific production measured by the number $n_{i}(t)$ of publications published  in year $t$, the cumulative number
  of citations $c_{i,p}(t)$ received by publication $p$, and our quantitative reputation measure defined here as the net citations aggregated across all publications $C_{i}(t)=\sum_{p}c_{i,p}(t)$.  \\

We begin with a description of our reputation model, followed by an empirical analysis of career trajectories, establishing $C_{i}$ as a good quantitative measure of reputation. We then establish quantitative  benchmarks  from the citation distribution within individual  publication portfolios and also 
 quantify features of the citation life-cycle, both of  which are crucial components of our reputation effect model. 
 Combining several empirical features of our analysis, we then investigate the role of the reputation effect,  showing that author reputation accounts for a significant 
fraction of the citation rate of young publications, thus providing a testable mechanism underlying cumulative advantage in science~\cite{Matthew1,cumadvprocess,BB2}.
 And finally, we develop a stochastic Monte Carlo reputation model which matches the  micro- and macroscopic citation dynamics.\\

\section*{Results}
\noindent{\bf Reputation signaling.} \
Academic career growth is a complex process emerging from the institutional, social,
and cognitive aspects of science. Conceptually, each career $i$ is embedded in two fundamental networks which are interconnected: the nodes in the first network represent scientists and in the second network represent publications.
The links within the first network represent collaborations between scientists, and within the second network they represent  citations between publications; the cross-links
represent the associations between individuals and their publications. 

Since these networks are dynamic, it is difficult to fully understand for any given individual, let alone the entire system, the 
complex information contained by all associations. As a result, scientific reputation has emerged as a key signaling mechanism to address the dilemma of excessive information
that arises, for example, in the task of evaluating, comparing, and ranking publication profiles in academic competitions. Reputation signals can flow between scientists $j\rightleftarrows k$, between publications $p \rightleftarrows q$,
and between a  publication and a scientist, $p \rightleftarrows i$. The focus of our analysis is on this latter dependency, $i\rightarrow p$, whereby author reputation  can impact the citation rate of his/her  publications,  generating  subsequent  reputation feedback, $p \rightarrow i$. 

Reputation plays an important role as a signal of trustworthiness and quality, a role  which addresses directly the ``agency problem'' characterizing the reward system in science \cite{StephanJEL}. Moreover, reputation signaling in scientific networks  is
used to overcome information asymmetries between scientists and other
academic agents; in this role it will become increasingly important as the rate of science publication grows and scientists have less time to absorb relevant advancements \cite{StephanJEL,OpenScience,SocialNetworksReputation,NetworkTiesReputation}.
With little time to read every paper on a given topic, this trustworthiness signal is 
anecdotally consistent with the common practice of perusing the author
names when  preliminarily evaluating the relevance of a
newly-found publication. In the past, an author's identity and associated reputation was mainly linked to reference lists and personal interactions. Nowadays, an author's reputation  is becoming increasingly visible through searchable publication databases, laboratory websites, press, and other media, in addition to citations.

\begin{figure*}
\centering{\includegraphics[width=0.75\textwidth]{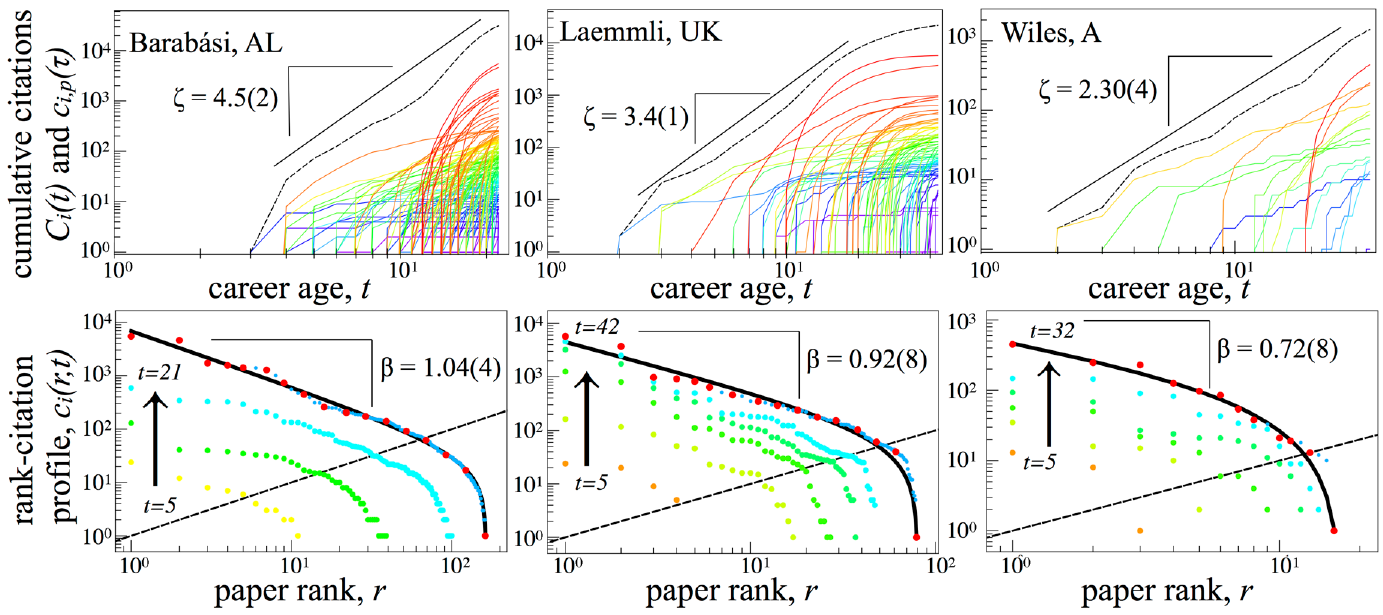}}
\caption{\label{empcareers} Quantitative patterns in the growth and size-distribution of the publication portfolio for scientists from 3 disciplines.  
(Left)  $c_{i,p}(t)$ for each author's most cited papers (colored according to net citations in 2010) along with
  $C_{i}(t)\sim t^{\zeta_{i}}$  (dashed black curve). (Right) The evolution of each author's rank-citation profile using
  snapshots taken at 5 year intervals. The darkest blue data points
  represent the most recent $c_{i}(r,t)$, and the subset of red data points indicate the logarithmically
  spaced data values used to fit the empirical data to our benchmark DGBD rank-citation distribution model \cite{RankCitSciRep} (solid black
  curve, see SI Appendix). The intersection of $c_{i}(r,t)$ with the dashed black line corresponds to the author's $h$-index $h_{i}(t)$.}
\end{figure*}

We measure the author reputation by $C_{i}(t)$, which measures not only the number of times his/her $N_{i}(t)$ publications have been referenced (an indication of overall scientific impact), but also the number of appearances of his/her name in the literature,  thereby providing a name-association visibility. What $C_{i}$ does not account for is intrinsic research quality, e.g. the quality ratio $C_{i}/N_{i}$ is broadly distributed across scientists. Since quantitative proxies for quality are limited to citation counts, it is presently difficult to distill the role played by quality in assessing overall scientific impact. 

By analyzing the top scientists, we reduce the 
compound reputation effect occurring when two or more scientists of comparable reputation  are  coauthors on a publication, a scenario where it may be
difficult to estimate the differential impact of these scientists on the citation rate. Due to data limitations requiring author name disambiguation and career data for all coauthors $j$,  we assume that a majority of the reputation signal is attributable 
to the central scientist $i$ by the approximation $C(t) \approx \sum_{j} C_{j}(t)  \approx C_{i}(t)$. Also, by analyzing top-cited cohorts, we can establish an upper bound to the strength of the reputation effect.
We note that $C_{i}$ possibly discounts the
role of mentor reputation effects early in the
career \cite{mentoreffect}. Nevertheless, by analyzing top scientists, the signaling
advantage  received early in their careers by associating with
prestigious mentors/coauthors should be negligible over the long run
\cite{SocialNetworksReputation}. 

To measure the role of author
reputation vis-\`a-vis publication impact, we use a regression model that
correlates the increase in the number of citations  $\Delta c_{i,p}(t+1)$ for a given paper $p$ in year $t+1$ using three explanatory variables: (i) the role played by the net number of citations $c_{p}(t)$ accrued up to paper age $\tau_{p}$ quantified by the power-law  regression parameter  $\pi$; 
 (ii) the role of publication age and the obsolescence of  knowledge quantified by the exponential regression parameter $\overline \tau$; and (iii) the role of author
reputation $C_{i}(t)$ quantified by the power-law  regression parameter $\rho$.

Together, these three features are (i) the publication citation effect $\Pi_{p}(t)\equiv[c_{p}(t)]^{\pi}$, (ii) the life cycle effect $A_{p}(\tau)\equiv\exp[-\tau_{p}/\overline\tau]$, and (iii) the author
reputation effect $R_{i}(t)\equiv[C_{i}(t)]^{\rho}$.
We perform a
multivariate regression to estimate the  $\pi$, $\overline \tau$, and $\rho$ values which parameterize the 
citation  model,
\begin{equation}
\Delta c_{i,p}(t+1) \equiv \eta  \times \Pi_{p}(t)  \times A_{p}(\tau) \times R_{i}(t)\ , 
\label{stochmodel}
\end{equation} 
 with the multiplicative log-normal noise term $\eta$. In the SI Appendix we  perform an additional fixed effects regression using  year as well as author variables to better control for the overall growth  in scientific output across time. In order to fully justify our  reputation effect model, in what follows, we first  account for two key features:    measures for cumulative career reputation and  obsolescence features of the citation life-cycle.\\

\begin{figure*}
\centering{\includegraphics[width=0.77\textwidth]{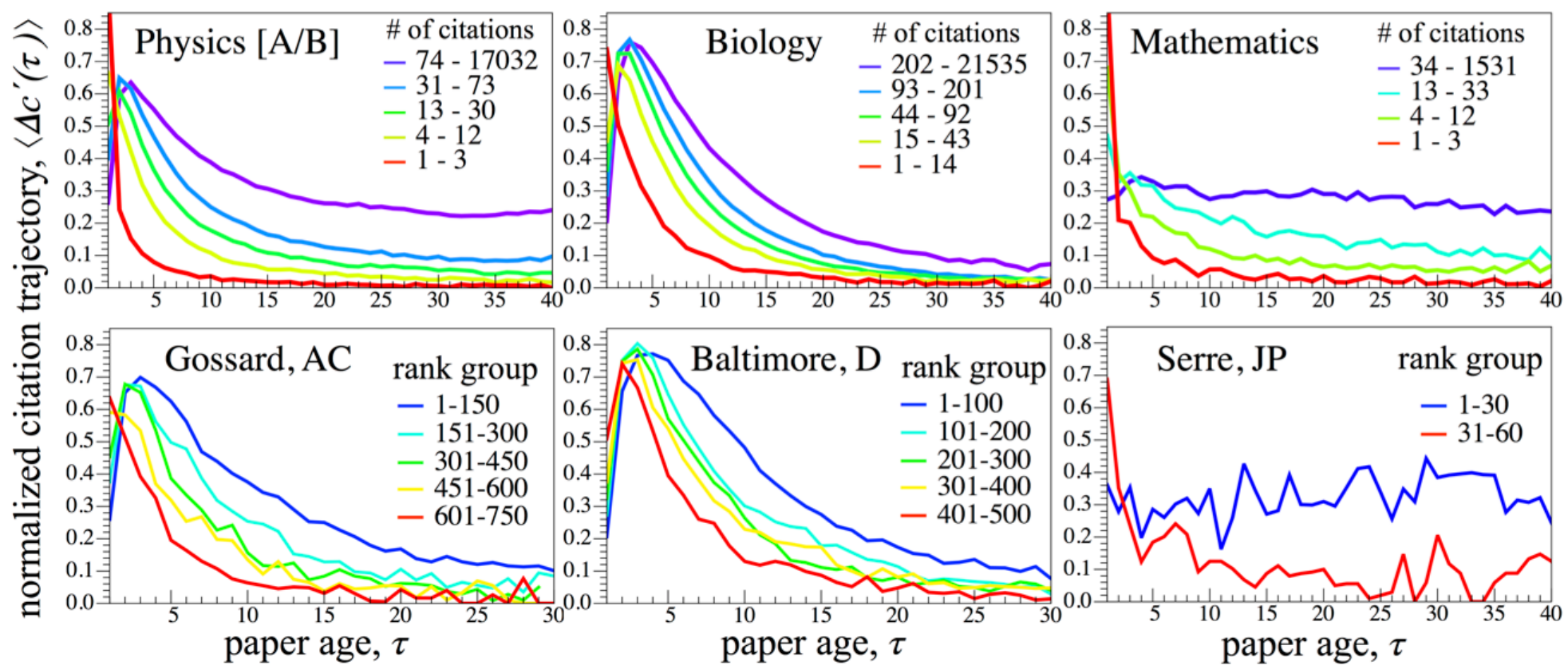}}
\caption{\label{lifecycle}  The citation life cycle  
  reflects both the intrinsic pace of discovery and the obsolescence rate of new knowledge, two features which are discipline dependent. (Left panels) For each of three disciplines, the averaged citation trajectory $\langle \Delta c'(\tau)
  \rangle$ is calculated for papers in the $n$-th quintile with the corresponding citation range indicated in each legend. For example, for physicists in dataset [A], the top 20\% of papers 
  have between 74 and 17,032 citations, and the papers in percentile 21--40 have between 31 and 73 citations. (Right panels) $\langle \Delta c'(\tau)
  \rangle$ calculated for
  rank-ordered groups of papers (listed in each legend) for 3 authors chosen from each discipline.}
\end{figure*}

\noindent{\bf Patterns of growth for longitudinal reputation measures.} \
In this section we investigate the patterns of cumulative publication and citation growth across the  career.
A striking statistical patterns observed for top scientists  is
the faster than linear growth in time, both in cumulative publication
number $N_{i}(t)\equiv\sum_{t'=1}^{t}n_{i}(t')$ and in cumulative
citation count $C_{i}(t)\equiv\sum_{p=1}^{N_{i}(t)}c_{i,p}(t)$ for a
large part of a scientist's ``growth phase,'' which we find to be
$\approx 30$ years after their first publication. 
Figures~\ref{NtCt}(A\&B) show the characteristic growth trajectories $\langle
N'(t) \rangle \sim t^{\overline\alpha}$ and $\langle C'(t) \rangle \sim
t^{\overline \zeta}$, calculated by an appropriate average over
individual $N_{i}(t)$ and $C_{i}(t)$, respectively. To facilitate visual comparison, we use arbitrary normalized
ordinate units so that each curve
starts from the same point,  $\langle
N'(1) \rangle = \langle
C'(1) \rangle \equiv 1$. The growth trajectories are
characterized by superlinear algebraic growth, with $\overline \alpha \gtrsim 1 $ and
$\overline \zeta > \overline \alpha$ (values shown in Fig.~\ref{NtCt}). 
Individual exponents
$\alpha_{i}$ and $\zeta_{i}$ are also calculated for the $N_{i}(t)$ and
$C_{i}(t)$ of each author, (in addition to multiple other quantitative measures, see SI Appendix, Tables S1--S9).  
We averaged both $\alpha_{i}$ and $\zeta_{i}$ within each dataset, 
confirming that $\langle \alpha _{i} \rangle \cong \overline \alpha$ and
$\langle \zeta _{i} \rangle \cong \overline \zeta$, confirming that the aggregate
patterns hold for the individual scale.
 In the SI Appendix we control for the exponential growth in scientific publication rates which can contribute to the longitudinal growth in $C_{i}(t)$. We define ``deflated'' citation counts $ \Delta c^{D}_{i,p}(t) \equiv \Delta c_{i,p}(t)/D(t)$ which are normalized  by the number of publications  $D(t)$  within a given discipline (since a new publication can cite an old publication only once). For each discipline we observe a  $5$\% exponential growth in $D(t)$ over the last half century. After deflating each $C_{i}(t)$, the net affect is only to reduce the estimated $\zeta_{i}$ values by roughly 15\%, meaning that the  growth  exponents $\zeta_{i} \gtrsim 2$ reflect significant growth above the underlying baseline growth trend in science.

Hence, we use $C_{i}(t)$ as a quantitative measure of reputation owing to the fact that the time dependence is readily quantified by a single parameter $\zeta_{i}$. We also use the power-law scaling  of $C_{i}(t)$ as a  benchmark for the stochastic career model we develop in the final section. 
Figure \ref{empcareers} shows two additional empirical benchmarks: (a) the microscopic citation dynamics of individual publications comprising  the publication portfolio and (b)  the rank-citation profile which is the Zipf distribution of the publications ranked in
decreasing order of rank $r$,  $c_{i}(1)\geq c_{i}(2)\geq\dots\geq c_{i}(N_{i})$. We confirm that the individual curves $c_{i}(r)$ belong to the class of the discrete generalized beta
distributions (DGBD), which in  the general form reads $c(r)\propto r^{-\beta}(N+1-r)^{\gamma}$ \cite{RankCitSciRep}. We validate the  DGBD fits using the $\chi^{2}$ test (see SI Appendix), also using $\beta_{i}$ and $\zeta_{i}$ as quantitative benchmarks for our MC model.\\

\noindent{\bf Variability in the citation life-cycle.} \
Important scientific discoveries can cause paradigm shifts and significantly boost the reputation of scientists associated with the discovery
\cite{citationboosts}.  In order to measure the reputation effect, one must also account for   obsolescence features of  scientific knowledge. It is also important to account for the variations in scientific impact, since most publications report results that are not seminal breakthroughs, but, rather, report incremental 
advances that are likely to have relatively short-term relevance.

In this section we analyze the dynamics of the citation
trajectory $\Delta c_{p}(\tau)$, the number of new citations received in
publication year $\tau$, where $\tau$ is the number of years since the
publication was first cited.  
We analyze $\Delta c_{p}(\tau)$ at two levels
of aggregation: (i) For each discipline, we calculate an averaged  $\Delta c_{p}(\tau)$   by collecting publications with similar total citation
  counts $c_{p}$. To achieve a scaled trajectory that is better suited for
  averaging we normalize each individual $\Delta c_{p}(\tau)$ by its
  peak citation value, $\Delta c_{p}'(\tau)\equiv\Delta c_{p}(\tau)/{\rm
    Max}[\Delta c_{p}(\tau)]$. In Fig.~\ref{lifecycle}, the panels on the left
  show the characteristic citation trajectory of publications belonging to
  each of the top 5 quintiles of each disciplinary citation distribution. Each
  curve represents the average trajectory $\langle \Delta c'(\tau)
  \rangle \equiv N_{q}^{-1} \sum_{p} \Delta c_{p}'(\tau)$ calculated
  from the $N_{q}$ publications in quintile $q$.
(ii) For each career $i$, we calculate $\langle \Delta c_{i}'(\tau) \rangle$ by averaging over groups of ranked citation
  sets within his/her publication portfolio.  The  panels on the right of Fig.~\ref{lifecycle}  show that even within prestigious careers, there is  significant variation in the citation life cycle.
  
  \begin{table*}
\centering{ {\small
\begin{tabular}{@{\vrule height .5 pt depth4pt  width0pt}l|ccc|cccc}
&\multicolumn3c{{\bf  $c(t-1) < c_{\times}$}}&\multicolumn3c{{\bf $c(t-1) \geq c_{\times}$}}\\
\noalign{
\vskip-1pt}
\hline
Name & $\pi_{i}$ (paper) & $\overline{\tau}_{i}$  (lifecycle) &  $\rho_{i}$ (reputation) & $\pi_{i}$ (paper) & $\overline{\tau}_{i}$ (lifecycle) & $\rho_{i}$ (reputation)   \\
\hline
\hline
Gossard, AC &	$ 0.34 \pm 0.027$ 	&	$ 4.92 \pm 0.261$ 	&	$ 0.25 \pm 0.008$ 	&	$ 0.80 \pm 0.048$ 	&	$ 4.73 \pm 0.184$ 	&	$ 0.09 \pm 0.024$  \\
Barab\'asi, AL	&	$ 0.42 \pm 0.036$ 	&	$ 3.00 \pm 0.155$ 	&	$ 0.29 \pm 0.010$ 	&	$ 1.06 \pm 0.016$ 	&	$ 3.65 \pm 0.111$ 	&	$ 0.01 \pm 0.011 $  \\
Ave. $\pm$ Std. Dev.	 [A] &	$ 0.43 \pm 0.14$ 	&	$ 5.67 \pm 2.52$ 	&	$ 0.22 \pm 0.06$ 	&	$ 0.96 \pm 0.19$ 	&	$ 8.93 \pm 4.09$ 	&	$ -0.07 \pm 0.11$ \\
\hline
\hline
Baltimore, D	&	$ 0.32 \pm 0.018$ 	&	$ 4.64 \pm 0.148$ 	&	$ 0.28 \pm 0.006$ 	&	$ 0.62 \pm 0.047$ 	&	$ 5.92 \pm 0.250$ 	&	$ 0.15 \pm 0.026$ \\
Laemmli, UK	&	$ 0.54 \pm 0.036$ 	&	$ 5.09 \pm 0.297$ 	&	$ 0.21 \pm 0.014$ &	$ 1.09 \pm 0.025$ 	&	$ 6.40 \pm 0.255$ 	&	$ -0.12 \pm 0.019$\\
Ave. $\pm$ Std. Dev.	 [D] &	$ 0.40 \pm 0.14$ 	&	$ 6.64 \pm 6.24$ 	&	$ 0.26 \pm 0.05$  & $ 0.99 \pm 0.22$ 	&	$ 9.55 \pm 26.30$ 	&	$ -0.06 \pm 0.14$  \\
\hline
\hline
Serre, JP	&	$ 0.33 \pm 0.095$ 	&	$ 15.90 \pm 3.724$ 	&	$ 0.14 \pm 0.026$ 	&	$ 0.66 \pm 0.065$ 	&	$ 20.50 \pm 3.862$ 	&	$ -0.03 \pm 0.039 $  \\
Wiles, A	&	$ 0.56 \pm 0.208$ 	&	$ 5.23 \pm 1.187$ 	&	$ 0.24 \pm 0.052$ 	&	$ 0.70 \pm 0.059$ 	&	$ 9.04 \pm 0.633$ 	&	$ 0.10 \pm 0.042$  \\
Ave. $\pm$ Std. Dev. [E]	&	$ 0.27 \pm 0.17$ 	&	$ 30.60 \pm 56.80$ 	&	$ 0.14 \pm 0.07$ &	$ 0.54 \pm 0.25$ 	&	$ 21.40 \pm 54.30$ 	&	$ 0.01 \pm 0.11$  \\
\hline
\end{tabular}}}
\label{table:regressALL}
\caption{Best-fit parameters for each effect ($\pm$ std. errors), both for individual  careers and the average values ($\pm$ std. dev.) calculated within each disciplinary dataset. The three features of the citation model are parameterized by  the publication citation effect ($\pi$),  the life-cycle effect ($\overline{\tau}$), and  the reputation effect ($\rho$). 
For statistical significances see SI Appendix Tables S10-S22.}
\end{table*}

At both levels of aggregation, the impact life cycle typically
peaks before publication age $\tau \approx 5$ years. Counterexamples likely correspond to publications
which receive a delayed secondary attention, e.g. receiving subsequent experimental validation of a previous theoretical prediction, and vice versa.  We
define the half-life $\tau_{1/2}$ as the time to reach half the peak
citation rate, $\Delta c'(\tau_{1/2})=1/2$ in the decay phase.  
Papers in the theoretical domains of mathematics and physics can  
exhibit $\tau_{1/2}>40$ years. Remarkably, some top mathematics publications even have $\tau_{1/2}$ that span nearly the entire data sample duration of $100$ years, reflecting the indisputable and foundational nature of ``progress by proof.''  This is in contrast to  top-cited  cell biology  publications, whereby for even the top 20\% of most cited works, the value $\tau_{1/2} \approx 10$ years. This relatively short decay timescale likely arises from the  large scale of research output in bio-medical fields, which leads to a significantly  higher discovery rate, and likewise, a relatively faster obsolescence rate.

The relation between the decay time scale $\overline \tau$ and $c_{p}$ provides insight into the knowledge diffusion rate. Fig. \ref{NtCt}(C) shows an approximate scaling relation $\tau_{1/2}\sim c_{p}^{\overline \Omega}$ when grouping publications 
into logarithmically spaced $c_{p}$ bins. Physics and biology differ mainly for the highly cited publications with $c_{p} \gtrsim 40$, whereas mathematics shows larger variation in $\tau_{1/2}$ per citation.
The $\overline \Omega$ value provides an approximate relation between citations and time. 
In mathematics $\tau_{1/2} \propto c_{p}$, indicating that the impact is distributed roughly uniformly across time. However, for biology publications the sub-linear relation with $\overline \Omega \approx 0.30$ indicates that for two publications, one with twice the citation impact as the other, the more cited publication gained twice the number of citations over a time period $\tau_{1/2}$ that was less than twice as large as the $\tau_{1/2}$ of the less-cited publication. The differences in $\overline \Omega$ are possibly related to discipline-dependent bursts in technological advancement,  funding initiatives \cite{EconScience}, and other social aspects of science that are related to non-linearities in scientific advancement.\\

\noindent{\bf Baseline citation model.} \
 To provide an initial test for basic mechanistic differences between the citation dynamics of
highly-cited publications and less-cited publications, in this subsection we analyze 
the relation between $\Delta c_{p}(t+1)$ and $c_{p}(t)$
representing the standard baseline preferential attachment (PA) model (corresponding to the limit $\overline \tau \rightarrow \infty$ and $\rho = 0$). 
Grouping together papers by $c_{p}(t)$  (using logarithmic bins), we  calculate for each group  the mean number of new citations in the following year, $\langle \Delta c_{p}(t+1) \rangle$.
Fig. \ref{NtCt}(D) shows the empirical relation for physicists in datasets [A/B], indicating that publications with citations above a gradual but substantial citation crossover value 
$c_{\times}$ obey a distinct scaling law that matches approximately linear ($\pi \approx 1$) preferential attachment dynamics (see SI Appendix, Fig. S8, for other disciplines).
However, below $c_{\times}$, the  citation rates are in excess of the citation rate expected from linear preferential attachment alone,
reflecting the citation premium that can be achieved via reputation. 

\begin{figure*}
\centering{\includegraphics[width=0.7\textwidth]{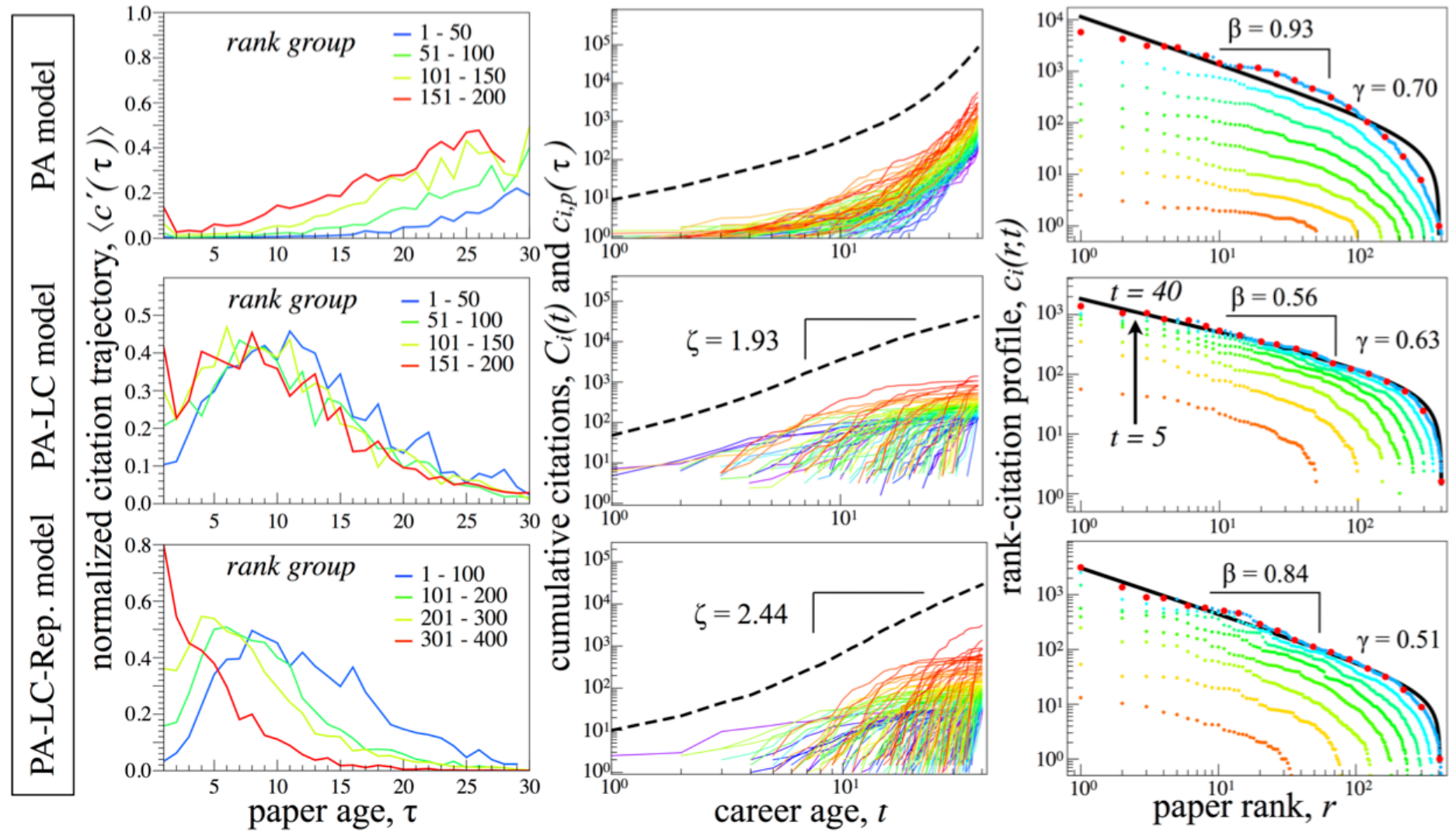}}
\caption{\label{MCcrmodel}  Comparison of three Monte Carlo career models against empirical benchmarks demonstrated in Figs. 1--3 and  S1--S3.
For each model we show $\langle \Delta c' (\tau) \rangle$ for the top 4 groups of ranked papers,  the evolution of $c_{i,p}(\tau)$ and $C_{i}(t)$ (dashed black curve), and the evolution of the rank-citation profile $c_{i}(r)$ at 5-period intervals. The best-fit DGBD $\beta$ and $\gamma$ parameters are also useful as quantitative benchmarks. 
For each model we evolve the system over $T\equiv 40$ periods, each period representative of a year. 
See SI Appendix for further elaboration of the model parameters used in the MC simulation.
}
\end{figure*}

\noindent{\bf Quantifying the role of the reputation effect.} \
 By analyzing the publications of highly-cited scientists, we have shown that the basic citation dynamics above and below the citation crossover value $c_{\times}$ vary considerably. 
In this subsection we investigate the role played by the reputation effect for  publications with $c_{p}(\tau) \geq c_{\times}$ compared to  
publications with $c_{p}(\tau) < c_{\times}$.  Based upon the  assessment of the growth dynamics (See SI Appendix, Figs. S8 and
S9), we choose the crossover values $c_{\times} \equiv 40$ [A/B],  $c_{\times} \equiv 10$ [C], $c_{\times} \equiv 100$ [D], and 
$c_{\times} \equiv 20$ [E]. Our results are not strongly dependent on reasonable variations around  our choice of $c_{\times}$. Table 1 shows the $\pi_{i}$, $\overline {\tau_{i}}$, and $\rho_{i}$  estimates,  above and below $c_{\times}$, for the individual careers highlighted in Figs. 1 and 3.  
For tables of the regression values aggregating over all careers in each disciplinary dataset see SI Appendix Tables S10--S13, and for the values for all 450 scientists analyzed individually see SI Appendix Tables S14 -- S22. 
The estimated model values are consistent when comparing between aggregated disciplinary datasets and individual career datasets.
Interestingly, we find that mathematicians exhibit relatively high life-cycle exponents $\overline {\tau_{i}}$ as compared to physicists and biologists, consistent with the empirical trajectories shown in Fig. \ref{lifecycle}.
However, the reputation effect  $\rho_{i}$ is less prominent in mathematics, possibly related to features of small team sizes and axiomatic discoveries which  may decrease the role of reputation effects in conveying prestige signals.

Our main result is a robust pattern of role switching by author- and publication-specific effects, specifically
\begin{equation}
 \rho(c< c_{\times}) >
\rho(c \geq c_{\times}) \ \text{ and } \ \pi(c< c_{\times}) < \pi(c \geq c_{\times}) \ .
\label{modelinequalities}
\end{equation} 
For example,  for the aggregate dataset [A/B] representing prolific physicists,  we estimate the values 
$\rho(c< 40) \approx 0.2$, $\rho(c \geq 40) \approx 0$, $\pi(c< 40) \approx 0.4$, and $ \pi(c \geq 40) \approx 1$.
To emphasize the role of reputation on new publications, consider two scientists separated by a factor of 10 in their cumulative citations,  $C_{1}(t)=10C_{2}(t)$. All other things being equal, the citation premium attributable to reputation alone for publications in the reputation regime  ($c<c_{\times}$) is  $\Delta c_{1}(t) / \Delta c_{2}(t) = 10^{\rho} \approx 1.66 $ (using the value $\rho = 0.22$ for dataset [A]).
Hence, there is a 66\% increase in the citation rate for each tenfold increase in  $C_{i}(t)$, which integrated over a career can provide significant positive feedback.
A pattern that emerges independent of discipline is  $\rho(c \geq c_{\times}) \approx 0$, meaning that reputation only plays a significant role for $c < c_{\times}$. 
 In the SI Appendix Section S6 we test the robustness of this result by implementing a fixed effects regression, the result of which reaffirms  the distinct roles of $\pi$ and $\rho$ above and below $c_{\times}$.
Hence, these two inequalities in Eq.~\ref{modelinequalities} indicate that publications are initially boosted by
author reputation to a citation   ``tipping point''  $c_{i,p} \approx c_{\times}$, above which the
citation rate is sustained in large by publication reputation.
These findings
show how microscopic  reputation  mechanisms contribute to cumulative
 ``rich-get-richer'' processes in science \cite{Matthew1,BB2}.

\noindent{\bf Simulating synthetic Monte Carlo careers with the reputation model.} \
Here we discuss three variants  of a Monte Carlo (MC) career growth model which simulates the dynamics of $\Delta c_{i,p}(t+1)$
for each publication $p$ in each time period $t$ of the career of synthetic author $i$.
With each variant we introduce progressively a new feature of publication citation trajectories. 
 (i) We begin with a basic linear preferential attachment model (PA model) whereby $\Delta c_{i,p}(t+1) \propto c_{i,p}(t)$. 
(ii) The PA-LC model includes a life-cycle (LC) obsolescence effect, $A_{p}(\tau)$. Fig. \ref{MCcrmodel} compares models (i-ii), which do not incorporate author specific factors,  with the reputation model (iii) given by Eq. \ref{stochmodel}. 
The PA model fails to reproduce the characteristic trajectories of real publications, since there is a clear first-mover advantage \cite{FirstMover} for the first publications published in the career, as well as non-power-law growth of $C_{i}(t)$. 

We use quantitative patterns demonstrated for real careers in Figs. \ref{empcareers}--\ref{lifecycle}  as empirical benchmarks to distinguish models (ii) and (iii). 
We confirm that the reputation model (iii) satisfies the empirical benchmark characteristics  in all 3 graphical categories (see Fig. \ref{MCcrmodel}).
We also confirm for the model (iii), but not for the model (ii), that there is a distinction between  $\langle \Delta c' (\tau_{p}) \rangle$ for different rank sets.
Furthermore, for model (iii) we quantitatively confirm that $C(t)\sim t^{\zeta}$ with
$2 \lesssim \zeta \lesssim 3$. For sufficiently large $t$  we also confirm that  
$c(r,t)$  belongs to the class of DGBD distributions, with $\beta$ values within the range of values observed empirically. In the SI Appendix text we further demonstrate how the model can be used to estimate properties of ``average'' careers for a given MC parameter set. For example, Fig. S11 shows excellent agreement between the reputation model's prediction and empirical data when estimating the fraction $f_{\geq c_{x}}$ of publications with $c_{p} \geq c_{\times}$ for a given career age $t$. Empirically, we observe saturation $f_{\geq c_{x}} \approx$ 0.20 to 0.30 for large $t$. 

\section*{Discussion}
 Social networks in science are characterized by
heterogeneous structure
\cite{StructuralHoles} that provides opportunities for intellectual and social
capital investment 
\cite{socialcapital} and  influences scientists' research strategies 
\cite{NetworkTiesReputation}.  
Identifying patterns of career growth is becoming increasingly important, largely due to the widespread emergence of quantitative  evaluation processes and recent efforts to develop quantitative models of career development. However, difficulties in accounting for complex social mechanisms, in addition to non-linearities and non-stationarities in the career growth process,   highlight the case for caution in the development of predictive career models  \cite{Caution1,Predictability2}.  Without a  better understanding of the institutional features and scientific norms that affect scientific careers, along the variable path from apprentice to group leader and mentor, there is a possibility to misuse  quantitative  career metrics  in the career evaluation process.

Toward the goal of  better understanding  career growth, with potential  policy implications for the  
quantitative career evaluation process,  we have analyzed the effect of reputation on the micro-level processes underlying the dynamics of a scientist's research impact. 
We used a regression model for the citation rate $\Delta c_{i,p}$ which accounts for the role of publication impact  ($\pi$), the role of knowledge obsolescence ($\overline \tau$), and the role of author reputation ($\rho$).
Interestingly, we find that the reputation parameter 
$\rho(c \geq c_{\times}) \approx 0$, meaning that in the long run the reputation effect makes a negligible contribution to the citation rate of papers with large $c_{p}$. However, we identify caveats concerning the way  publications can become highly cited.
By analyzing the variation of $\rho$ and $\pi$ for publications above and below a citation threshold $c_{\times}$ 
we identify the advantageous role that author reputation plays in the citation dynamics of new publications, finding  that future publications can gain roughly a 66\% increase in  $\Delta c$ for each tenfold increase in reputation $C_{i}$.
 We note that it is also likely that both institutional affiliation and journal reputation also play a role in the citation dynamics, however disentangling the interaction between the multiple reputation sources will likely be challenging and remains an open avenue for investigation. 

In the process of analyzing the effect of reputation on career growth, it was necessary to also quantify two essential features of our model, namely patterns of cumulative productivity and impact across the career, and patterns of obsolescence in the citation life cycle of individual publications. 
For prolific scientists, we have identified a robust pattern of growth for two cumulative reputation measures, $N_{i}(t)$ and $C_{i}(t)$, each of which are quantifiable by a single scaling parameter, $\alpha_{i}$ and $\zeta_{i}$, respectively. These  regularities  suggest that underlying  social
processes  sustain career growth via reinforcing  coevolution of scientific collaboration and publication 
\cite{GrowthCareers,Borner,TeamAssembly,socialgroupevol}. 
We also introduced a citation deflator index to control for the increased supply of citations arising from the exponential  5\%  growth (per year) in the total publication output. Analyzing the growth of 'deflated' citation trajectories, $C^{D}_{i}(t)$, we observed $\zeta_{i} \gtrsim 2$ values which confirms that the observed career growth is significantly above the baseline inflation rate of science.
We  note that in  using non-decreasing cumulative reputation measure $C_{i}(t)$, we have overlooked the possibility that reputation can significantly decrease, as occurs when a scientist is associated with invalidated and/or fraudulent science. Indeed, recent evidence indicates that  the retraction of a publication can have a negative impact on the potential growth of $C_{i}$ \cite{RetractionCost}. As a robustness check  we also used the annual citation rate $\Delta C_{i}(t)$ as an additional (non-cumulative) reputation measure, one  that is more amenable to controlling for secular growth trends. We applied a multivariate fixed effects regression using $\Delta C_{i}(t)$ as the reputation measure (see SI Appendix Section S6), which reconfirms  the role of reputation in citation dynamics.


Our analysis tracks the evolution of each scientist's publication portfolio across the career, suitably illustrated by the rank-citation profile $c_{i}({r})$,  which highlights the skewed distribution of  $c_{i,p}$, even within a career. Arising from the power-law features of  $c_{i}({r})$ \cite{RankCitSciRep}, we emphasize the disproportionate fraction of a scientist's total citations $C_{i}$ owed to the $c_{i}(r=1)$ citations coming from his/her highest-cited publication. For example, the average and standard deviation of the ratio $c_{i}(1)/C_{i}$  is $0.15 \pm 0.13$ for the  physicists, $0.09 \pm 0.08$ for the  biologists, and $0.16 \pm 0.08$ for the  mathematicians we analyzed, which emphasizes the potentially large reputation boost that can follow from just a single high-impact publication.  With rapidly increasing numbers of journals accompanied by the opportunity for rapid publication, the reputation  effect   provides an  incentive to aim for  quality over quantity in the publication  process, reinforcing a research strategy which is beneficial for science and scientists.

It  is also important to consider the role of reputation in light of the increasing orientation of science around team endeavors characterized by multiple levels of hierarchy and division of labor \cite{TeamScience}. 
Because it is difficult to evaluate and assign credit to  individual contributions in a team setting, 
there may be an increase in the role and strength of the reputation in overcoming the problem associated with asymmetric and incomplete information. 
In addition to the collaboration network, reputation also plays a key role in numerous other   scientific inputs (money, labor, knowledge, etc.) which inevitably affect the overall quantity and quality of scientific outputs. It will become increasingly important to understand the relation between these  inputs and outputs  in order to efficiently allocate scientific resources  \cite{GrowthCareers,EconScience,ResourcesScience}. 

In light of individual careers, an institutional setting based on quantitative appraisal that neglects these complex relations may inadvertently go against the goal of sustaining the
careers of talented and diligent young academics \cite{GrowthCareers}. 
For example, our finding of a crossover behavior around $c_{\times}$ shows how young scientists
lacking reputation can be negatively affected by social stratification
in science. The appealing competitive advantage gained by working with a prestigious mentor may be countered by the possibility that it may not be the ideal mentor-advisee match. Despite having analyzed cohorts of  highly cited scientists, our results have broad implications across the scientific population when one considers the numerous careers that interact with top scientists via  collaboration or mentorship. 

In excess, the reputation effect may also negatively affect science, especially considering how online visibility has become a relatively new reputation platform in an increasingly competitive environment. As such, strategies of self-promotion may emerge as scientists try to ``game''  with reputation systems.  In such scenarios, it may  be hard to disentangle fair from foul play.  For example, it may be difficult to distinguish self-citation strategies  aimed at boosting $C_{i}$ from the natural tendency for  scientists who are crossing disciplinary borders to self-cite with the intention to send credibility signals
 \cite{SelfCiteScharnhorst}. Reputation will also become increasingly important in light of preferential treatment in search queries, e.g. Google Scholar, which provide query results ordered according to citation measures. These systemic  search and retrieval features may further strengthen association of reputation  between publications and authors. In all, our results should motivate future research to inspire institutional and funding body evaluation schemes to appropriately
account for the roles that reputation and social context play in science. For example, our results can be used in support of the double-blind review system, which by reducing the role of reputation, is perceived to have advantages due to its  objectivity and fairness \cite{PeerReview}.  
We conclude with a general note that the data deluge brought forth during the past decade is fueling extensive efforts in the computational social sciences\cite{CompSocScience} 
to identify and study the so-called ``social atom'' \cite{SocialAtom}. Because our methodology is general, we speculate that other social networks  characterized by trust and partial/asymmetric information are also based on similar reputation mechanisms. Indeed, it is likely that agent-based reputation mechanisms will
play an increasing role due to the omnipresence of online recommender
systems governed by reputation dynamics operating as a general diffusive contagion phenomena \cite{VespignaniSocioTechnical2}.\\

\bigskip
\bigskip

\noindent{\bf Supporting Information (SI) Appendix} available at  \href{http://physics.bu.edu/~amp17/webpage_files/MyPapers/Reputation_SI.pdf}{AMP homepage}\\

\noindent{\bf Acknowledgements} We thank M. C. Buchanan, A. Scharnhorst, J. N. Tenenbaum, S. V. Buldyrev, V. Tortolini, A. Morescalchi, and
  S. Succi for helpful discussions and each referee for their extremely valuable comments and insights. AMP acknowledges  COST Action MP0801 and COST Action TD1210.  AMP, SF, KK, MR, and FP thank the
  EU FP project ``Multiplex'' and AMP, MR, and FP acknowledge PNR ``Crisis
  Lab'' project at IMT. OP acknowledges funding from the Canadian SSHRC.
  HES thanks NSF grant CMMI 1125290.

\end{document}